\pdfoutput=1  % arXiv: select pdflatex. Must stay within the first 5 lines.
\documentclass[11pt]{article}

\usepackage[utf8]{inputenc}
\usepackage[T1]{fontenc}
\usepackage{lmodern}
\usepackage[margin=1in]{geometry}
\usepackage{amsmath,amssymb}
\usepackage{array}
\usepackage{booktabs}
\usepackage{longtable}
\usepackage{graphicx}
\usepackage{xcolor}
\usepackage{tikz}
\usepackage{microtype}
\usepackage[numbers,square,comma,sort&compress]{natbib}
\usepackage{hyperref}
\usepackage{url}
\hypersetup{
  colorlinks=true,
  linkcolor=blue!55!black,
  citecolor=blue!55!black,
  urlcolor=blue!55!black,
  pdftitle={Is INT8 Portable? A Cross-Platform Measurement Study of Quantized Inference on Embedded and Automotive Accelerators},
  pdfauthor={Yuyeong Shin}
}

\newcolumntype{L}[1]{>{\raggedright\arraybackslash}p{#1}}
\newcolumntype{C}[1]{>{\centering\arraybackslash}p{#1}}
\newcolumntype{R}[1]{>{\raggedleft\arraybackslash}p{#1}}

\title{Is INT8 Portable?\\[2pt]
\large A Cross-Platform Measurement Study of Quantized Inference on Embedded and Automotive Accelerators}

\author{Yuyeong Shin\thanks{Contact: \texttt{yyshin@katech.re.kr}. Korea Automotive Technology Institute (KATECH).}}

\date{}

\begin{document}
\maketitle

% ---------------------------------------------------------------------
\begin{abstract}
Eight-bit integer (INT8) post-training quantization is the default recipe for edge deployment, under a widely held assumption: INT8 makes inference faster at a small, predictable accuracy cost, and a model quantized once can be carried to any target. We test that assumption with a controlled measurement study across seven hardware classes --- ARM and x86 CPUs, a discrete GPU, an NVIDIA Jetson AGX Orin iGPU and its NVDLA cores, and two vendor NPUs (Qualcomm Hexagon HTP, DEEPX DX-M1) --- holding the ONNX artifact and the quantization scales fixed so the integer kernel or ISA is the only free variable. Portability fails on three axes. \textbf{(1)}~The \emph{sign} of the INT8 speedup is set by the CPU's dot-product ISA (ARM \texttt{dotprod}/SDOT, x86 VNNI): cores that have it speed up by up to $2.1\times$, cores that lack it slow \emph{down} by $1.7\times$, for the identical model and runtime. \textbf{(2)}~INT8 outputs are not portable, and the rule is an invariance rather than a gradient: FP32 predictions are bit-identical for every pair (1000/1000), while INT8 predictions agree 1000/1000 exactly when two targets share an integer kernel and 958--965/1000 whenever they do not --- independent of whether the boundary is CPU$\leftrightarrow$CPU or CPU$\leftrightarrow$accelerator, and invisible to top-1 accuracy, which is preserved. \textbf{(3)}~Vendor NPUs own quantization: a bring-your-own QDQ graph fails silently on one NPU (external scales ignored, accuracy $0.75 \rightarrow 0.005$ while it compiles, profiles and runs without error) and loudly on the other (the compiler refuses the graph), so only the vendor's native path yields a correct engine. We further show that edge-NPU latency \emph{regimes} are set by output/device-to-host transfer size rather than compute, and locate the transition with a fixed-compute sweep. We release the scripts and 32 reports. ``Quantize once, deploy anywhere'' is unsafe for embedded and automotive deployment, where per-input determinism and redundancy matter.
\end{abstract}

% ---------------------------------------------------------------------
\section{Introduction}

Quantizing a floating-point network to 8-bit integers is the first and often only compression step in an embedded deployment pipeline. The operational folklore is compact and appealing: \emph{(i)}~INT8 is faster than FP32/FP16 because integer math is cheaper and moves less memory; \emph{(ii)}~the accuracy cost is small and bounded; and \emph{(iii)}~a model quantized once is a portable artifact --- an ONNX file with QDQ nodes, or a TFLite model --- that can be handed to any runtime or accelerator. This folklore underlies the way quantization is taught, benchmarked (report one speedup and one accuracy number), and shipped.

This paper asks whether the folklore survives contact with real, heterogeneous hardware. We ran a controlled measurement campaign across seven hardware classes and three model families (image classification, CNN object detection, and transformer detection), taking care to hold the model artifact and --- critically --- the quantization \emph{scales} fixed across targets so that any difference we observe is attributable to the target's integer kernel or instruction set, not to a different quantizer.

We find that all three parts of the folklore fail, and they fail in ways that matter specifically for embedded and automotive systems:

\begin{enumerate}
\item \textbf{The speedup can be negative, and its sign is an ISA property (\S4).} For the identical INT8 model on the identical CPU runtime (ONNX Runtime, MLAS) a Raspberry Pi 5 Cortex-A76 and a Jetson A78AE --- which have ARM dot-product instructions --- get $1.83\times$ and $2.11\times$ \emph{faster}, while an x86 Core i9 without AVX-512 VNNI and a Cortex-A53 without dot-product get $1.65$--$1.76\times$ \emph{slower}. INT8 is not ``faster''; it is faster \emph{iff} the target has the right accumulate instruction.

\item \textbf{INT8 outputs are not numerically portable (\S5, headline).} FP32 predictions are bit-identical across every platform pair where FP32 was measured (1000/1000). INT8 predictions are not: they disagree on ${\sim}4$\% of inputs across a CPU$\leftrightarrow$CPU pair and across the CPU$\leftrightarrow$GPU boundary (built from the \emph{same} QDQ scales), and the rule is an invariance rather than a gradient: agreement is 1000/1000 whenever two targets share an integer kernel and 958--965/1000 whenever they do not, independent of whether the boundary is CPU$\leftrightarrow$CPU or CPU$\leftrightarrow$accelerator. A CPU$\leftrightarrow$vendor-NPU pair diverges further (939/1000), but its scales cannot be held fixed, so we report it as a deployment observation outside the invariance claim. Top-1 accuracy is preserved --- the flips are net-neutral --- so this is invisible to a standard accuracy report, yet it means a quantized model is not a deterministic function of its input once you change the target. Concurrent work localizes the underlying epilogue-rounding mechanism by swapping INT8 kernels on a \emph{single} GPU for LLMs \citep{chen2026integeralibi}; our contribution is orthogonal and, for deployment, more consequential --- the cross-\emph{physical-device} measurement for vision/detection models under an identical ONNX file and identical scales, with FP32 bit-identity as a control (\S5).

\item \textbf{Vendors own quantization (\S6).} A model you quantized yourself is not deployable to a vendor NPU as-is. Qualcomm's Hexagon HTP \emph{silently} ignores external QDQ scales and collapses accuracy ($0.75 \rightarrow 0.005$) while compiling, profiling, and running without error; the DEEPX compiler \emph{loudly} rejects the same class of graph. Both are correct only through the vendor's own quantization path. These are opposite symptoms of one fact: the accelerator, not your toolchain, defines the numerics.
\end{enumerate}

Beyond portability, we contribute a systems observation that reframes edge-NPU performance analysis: \textbf{latency regime is set by output/data-movement size, not compute (\S7).} On one M.2 NPU, a classifier with a 4\,KB output is compute-bound and scales $2.19\times$ across cores, while a detector with a 2.82\,MB raw output on the same device, same runtime, is data-movement-bound and does not scale at all --- and a \emph{lighter}-compute detector with an even larger output likewise refuses to scale ($1.02\times$) and runs $26\times$ slower than the classifier. A controlled single-variable sweep --- compute held fixed, output swept $1020\times$ --- then traces the full transition curve and shows the regime boundary is not a knife-edge but a band whose width equals the core count ($N$ cores share one data-movement link), yielding a closed-form rule for the output size at which adding cores stops helping. We isolate output size as the causal variable.

We frame these findings for their intended audience. The submitting institution works on automotive edge AI, where the failures above are not academic: non-portable INT8 numerics undermine the determinism and cross-module consistency that safety cases and redundant (dual-compute) architectures rely on. \S8 and \S9 add supporting characterization (transformer INT8 breakdown, DLA behavior) and a catalog of silent-failure pitfalls we hit, and \S10 states the study's limits honestly.

\paragraph{Contributions.}
\begin{itemize}
\item A controlled, same-artifact/same-scale cross-platform methodology that isolates the integer kernel/ISA as the only variable (\S3).
\item C1: the INT8 speedup sign is determined by the CPU dot-product ISA, shown across four CPUs (\S4).
\item C2 (headline): INT8 numerical non-portability across CPU$\leftrightarrow$CPU and CPU$\leftrightarrow$accelerator boundaries, with FP32 as a bit-identical control (\S5).
\item C3: vendor NPUs own quantization; two opposite BYO-QDQ failure modes (\S6).
\item C4: edge-NPU bottleneck regimes are set by output/data-movement size, not compute; a controlled fixed-compute sweep traces the transition, whose boundary is a band of width $=$ core count (\S7).
\item A reproducibility artifact: scripts plus 32 measurement reports, and a claim-to-artifact map (Appendix~\ref{app:map}).
\end{itemize}

% ---------------------------------------------------------------------
\section{Background and Related Work}

\paragraph{Quantization foundations.}
Integer-arithmetic-only inference --- mapping weights and activations to INT8, accumulating products in INT32, and re-scaling back --- was formalized by Jacob et al.\ \citep{jacob2018} and codified for post-training use in the whitepapers of Krishnamoorthi \citep{krishnamoorthi2018} and Nagel et al.\ \citep{nagel2021whitepaper}, with a broad survey by Gholami et al.\ \citep{gholami2021survey}. The per-tensor vs.\ per-channel and symmetric vs.\ asymmetric design axes, and the result that symmetric INT8 typically stays within ${\sim}1$\% of FP32 for CNNs, are established by NVIDIA's evaluation \citep{wu2020}; rounding choice at quantization time is itself accuracy-critical \citep{nagel2020adaround}. Our work proposes no new quantization algorithm --- it measures how \emph{existing}, correctly-produced INT8 behaves once it crosses a hardware boundary. Crucially, Jacob's pipeline makes the INT32 accumulate exact and order-independent; the re-scaling \emph{epilogue} is where implementation freedom --- and, as we and concurrent work show, non-portability --- lives.

\paragraph{Integer kernels and dot-product ISAs.}
The INT8 speed advantage depends on hardware dot-product/accumulate instructions: ARM \texttt{dotprod} (SDOT/UDOT, ARMv8.2-A) \citep{armisa} and x86 AVX-512 VNNI \citep{intelisa}, exploited by the low-precision GEMM libraries gemmlowp \citep{gemmlowp}, XNNPACK \citep{xnnpack}, FBGEMM \citep{khudia2021fbgemm}, and Microsoft's MLAS \citep{mlas} (the kernel library our ONNX Runtime CPU path uses). Prior characterization of data-center INT8 inference \citep{park2018facebook} documents this dependence for throughput. We contribute the \emph{cross-device sign-flip} framing --- that the same model and runtime is faster or slower depending only on whether the target CPU has these instructions --- and connect it to the accuracy-side consequence (\S4 $\rightarrow$ \S5).

\paragraph{Edge and mobile inference benchmarking.}
MLPerf Inference \citep{reddi2020mlperf}, MLPerf Tiny \citep{banbury2021mlperftiny}, MLPerf Mobile \citep{reddi2020mobile}, and AI-Benchmark \citep{ignatov2018, ignatov2019} are the standard cross-device benchmarks, and recent studies benchmark detectors across the exact hardware family we use (Jetson, Raspberry Pi 5, Coral) \citep{edgedetection2024, millar2025}. By design these quantize \emph{per submission and per backend} and report per-device throughput/accuracy scores; none holds the quantization scales fixed across targets, and none reports cross-target numerical \emph{agreement}. That axis --- numerical portability under fixed scales --- is our C2.

\paragraph{Numerical reproducibility and determinism (closest to our headline).}
The mechanism behind C2 was, concurrently with this work, localized by Chen \citep{chen2026integeralibi, chen2026deterministic}: swapping only the INT8 GEMM kernel (CUTLASS vs.\ Triton) inside an LLM serving stack \emph{on a single GPU} yields two engines that are each bit-reproducible against themselves yet agree on no generated sequence, with the divergence traced to scale application and output rounding in the \emph{epilogue} (the INT32 accumulate being exact), and power-of-two scales restoring bit-identical cross-kernel agreement. We cite this as concurrent prior work and claim no discovery of the mechanism. Our contribution is orthogonal in setting and method: Chen swaps kernels on one device for LLMs, whereas we measure per-input prediction disagreement \emph{across physical hardware boundaries} --- dot-product CPU $\leftrightarrow$ non-dot-product CPU, CPU $\leftrightarrow$ GPU, and CPU $\leftrightarrow$ vendor NPU --- for vision and detection models, using an identical ONNX file with identical embedded QDQ scales, and we contrast the INT8 divergence against FP32 bit-identity on those same devices. More broadly, MQBench \citep{li2021mqbench} measures a hardware-deployability \emph{accuracy gap} across backends but not per-input, bit-level disagreement under identical scales; the inference backend has been shown to confound even greedy-decoding LLM behavior \citep{masoudian2026}; and floating-point non-associativity \citep{fpnonassoc2024} and fixed-reduction-order remedies \citep{repdl2025} frame the FP side --- we scope our ``FP32 bit-identical'' claim to our observed, fixed-thread configuration accordingly (\S5, \S10). The divergence is folklore in practitioner issue trackers for quantized TFLite CPU-vs-NPU and cross-EP ONNX Runtime outputs, but to our knowledge has not been systematically measured across embedded and automotive accelerators.

\paragraph{Transformer quantization.}
Transformer INT8 fragility is driven by activation outliers \citep{dettmers2022llmint8, bondarenko2021}, addressed by activation migration (SmoothQuant \citep{xiao2022smoothquant}), activation-aware weight scaling (AWQ \citep{lin2023awq}), and weight-only PTQ (GPTQ \citep{frantar2022gptq}); for vision transformers specifically, PTQ4ViT \citep{yuan2022ptq4vit} and Liu et al.\ \citep{liu2021ptqvit} handle post-softmax/GELU activations. We use these to explain \emph{why} DETR INT8 collapses on-device and to quantify how much the activation-granularity lever recovers on a real toolchain vs.\ in fake-quant (\S8) --- reinforcing that activations, not op selection, are the fragile axis.

\paragraph{Vendor NPU toolchains.}
Vendor compilers differ in scaling, clipping, and kernel support, so the same checkpoint yields inconsistent cross-backend accuracy --- a point made by MQBench \citep{li2021mqbench} and, most directly, by Quant-Trim \citep{dhahri2025quanttrim}, which proposes a training-time hardware-neutral checkpoint. Qualcomm's QNN/AI-Hub stack quantizes to its own native format \citep{qualcomm_qnn}, as do Apple Core ML \citep{coreml} and TFLite/LiteRT delegates \citep{litert}. These works establish that vendors prefer their own quantization; we contribute a controlled, cross-vendor account of \emph{bring-your-own-QDQ rejection} and its two opposite failure modes --- silent (accuracy collapse while running) vs.\ loud (compile refusal) --- and show the native path is both correct and faster (\S6).

\paragraph{Accelerator characterization.}
The compute-bound vs.\ memory-bound dichotomy is the Roofline model \citep{williams2009roofline}; the primacy of data movement over compute energy is the Eyeriss line of work \citep{chen2016eyeriss, sze2017efficient}. We extend the deployment-level picture with a third, output-size-driven regime --- device-to-host (D2H)/PCIe-bound --- and show that on one PCIe-attached edge NPU the bottleneck (and whether multi-core helps) is set by the model's output tensor size, not its compute (\S7). Concurrent-inference profiling on Jetson \citep{jetsonconcurrent2025} corroborates our related finding that GPU-fallback subgraphs serialize otherwise-parallel accelerator work.

\paragraph{NVDLA and fixed-function INT8 accelerators.}
The Orin DLA is an NVDLA v2 instance \citep{nvdla, farshchi2019nvdla}; we characterize it as an INT8-only, CNN-favoring datapath that is the perf-per-watt leader for CNNs but fragments on transformers (\S8).

\paragraph{Automotive compute and redundancy (framing).}
Redundant, diverse compute across CPU/GPU/DLA is the backbone of automotive functional-safety architectures (ISO~26262 / ASIL \citep{iso26262}, NVIDIA DRIVE \citep{nvidiadrive}) and heterogeneous AV-SoC scheduling \citep{hetsched2022}. This is our motivation: non-portable INT8 numerics (\S5) directly threaten the cross-module agreement that dual-compute redundancy assumes --- the concern that motivates our planned follow-on work on a multi-module platform (\S11).

% ---------------------------------------------------------------------
\section{Experimental Methodology}

\subsection{Hardware matrix}
Table~\ref{tab:hw} lists the eight targets, grouped into the seven hardware classes this study spans, and the role each one plays.

\begin{table}[htbp]
\centering
\small
\begin{tabular}{L{2.7cm} L{5.0cm} L{4.8cm}}
\toprule
\textbf{Class} & \textbf{Target} & \textbf{Role in the study} \\
\midrule
ARM CPU (dotprod)      & Raspberry Pi 5, Cortex-A76                 & C1 sign, C2 agreement \\
ARM CPU (dotprod)      & Jetson AGX Orin, Cortex-A78AE              & C1 sign, C2 CPU$\leftrightarrow$accelerator \\
ARM CPU (no dotprod)   & i.MX8M-Nano, Cortex-A53                    & C1 sign (negative) \\
x86 CPU (no VNNI)      & Core i9-10900K                            & C1 sign (negative), C2 \\
Discrete GPU           & RTX 3080 (Ampere)                         & precision ladder, transformer INT8 \\
Edge iGPU + NVDLA      & Jetson AGX Orin (iGPU, $2\times$NVDLA v2) & accelerator char., C2 GPU kernel \\
Vendor NPU (mobile/auto) & Qualcomm Hexagon HTP (QCS8550, SA8775P) & C3 (silent BYO-QDQ) \\
Vendor NPU (auto)      & DEEPX DX-M1 (M.2, PCIe Gen2$\times$1)     & C3 (loud BYO-QDQ), C4 regimes \\
\bottomrule
\end{tabular}
\caption{Hardware matrix. Every device is a single unit ($n=1$ per class); we therefore make \emph{relative} claims about representative devices, not population claims about an ISA (see \S10).}
\label{tab:hw}
\end{table}

\subsection{Models and datasets}
ResNet-18/50 (ImageNet-1k classification), DETR-ResNet-50 (COCO detection, transformer), YOLO26n and YOLOv5s (COCO detection, CNN), plus BEVFormer/BEVDet (3D BEV) used only for latency/engine characterization. Accuracy is reported on ImageNet val (subset or full, stated per result) and COCO val2017 subsets. Absolute accuracy/latency are not comparable across sections because batch size, input resolution, and evaluation subset differ; we report \emph{relative} deltas within a controlled comparison.

\subsection{Controlled-comparison principle}
The core methodological device of this paper: for any cross-target comparison, we fix the model artifact and the quantization \emph{scales}, so the only free variable is the target's integer kernel/ISA. Concretely, the same \texttt{resnet50\_int8\_qdq.onnx} (with its embedded QDQ scales) is (a)~run on multiple CPU EPs and (b)~used to \emph{build} the TensorRT INT8 engine --- so the GPU integer kernel and the CPU integer kernel consume identical scales. When a comparison cannot hold scales fixed (e.g., a vendor NPU that rejects external QDQ), we say so and treat the result as a deployment finding, not a kernel comparison.

\subsection{Measurement protocol and scope}
Latencies are event-timed on GPU/accelerator paths and wall-clock on CPU/harness paths (the two are not directly comparable and are never mixed within a claim). Unless noted, batch size is 1. Numerical agreement is reported as the number of inputs (out of a fixed bundle, typically 1000 for classification) on which two targets produce the \emph{same} top-1 prediction. \textbf{Known limitation:} most latencies are single-run p50 without confidence intervals, and several accuracy numbers are on subsets; \S10 quantifies why this bounds our claims to relative comparisons, and this study's own \S9 shows subset evaluation can inflate top-1 by ${\sim}9.77$ percentage points.

% ---------------------------------------------------------------------
\section{The Speedup Sign Is ISA-Determined (C1)}

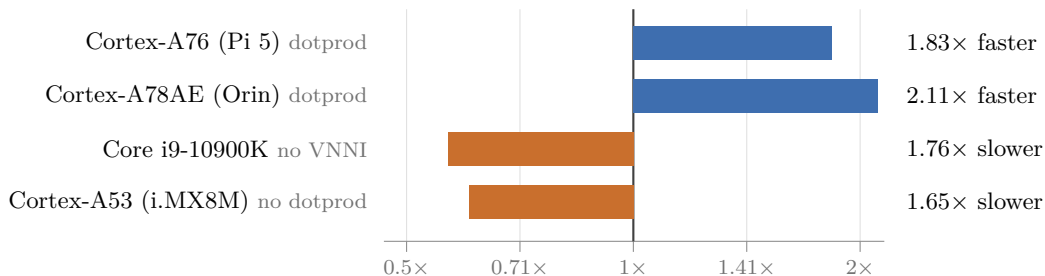
\begin{figure}[htbp]
\centering
\footnotesize
\begin{tikzpicture}[x=1cm,y=1cm]
\definecolor{cfast}{RGB}{62,110,175}
\definecolor{cslow}{RGB}{201,111,45}
\draw[black!12] (-3.000,0.30) -- (-3.000,-2.770);
\draw[black!45] (-3.000,-2.770) -- (-3.000,-2.910);
\node[below,black!55,font=\scriptsize] at (-3.000,-2.910) {0.5$\times$};
\draw[black!12] (-1.500,0.30) -- (-1.500,-2.770);
\draw[black!45] (-1.500,-2.770) -- (-1.500,-2.910);
\node[below,black!55,font=\scriptsize] at (-1.500,-2.910) {0.71$\times$};
\draw[black!12] (0.000,0.30) -- (0.000,-2.770);
\draw[black!45] (0.000,-2.770) -- (0.000,-2.910);
\node[below,black!55,font=\scriptsize] at (0.000,-2.910) {1$\times$};
\draw[black!12] (1.500,0.30) -- (1.500,-2.770);
\draw[black!45] (1.500,-2.770) -- (1.500,-2.910);
\node[below,black!55,font=\scriptsize] at (1.500,-2.910) {1.41$\times$};
\draw[black!12] (3.000,0.30) -- (3.000,-2.770);
\draw[black!45] (3.000,-2.770) -- (3.000,-2.910);
\node[below,black!55,font=\scriptsize] at (3.000,-2.910) {2$\times$};
\draw[black!45] (-3.300,-2.770) -- (3.320,-2.770);
\draw[black!75,thick] (0,0.30) -- (0,-2.770);
\fill[cfast] (0.000,-0.370) rectangle (2.622,0.070);
\node[left,font=\footnotesize] at (-3.400,-0.150) {Cortex-A76 (Pi~5)~\textcolor{black!55}{\scriptsize dotprod}};
\node[right,font=\footnotesize] at (3.500,-0.150) {1.83$\times$ faster};
\fill[cfast] (0.000,-1.070) rectangle (3.235,-0.630);
\node[left,font=\footnotesize] at (-3.400,-0.850) {Cortex-A78AE (Orin)~\textcolor{black!55}{\scriptsize dotprod}};
\node[right,font=\footnotesize] at (3.500,-0.850) {2.11$\times$ faster};
\fill[cslow] (-2.450,-1.770) rectangle (0.000,-1.330);
\node[left,font=\footnotesize] at (-3.400,-1.550) {Core i9-10900K~\textcolor{black!55}{\scriptsize no VNNI}};
\node[right,font=\footnotesize] at (3.500,-1.550) {1.76$\times$ slower};
\fill[cslow] (-2.170,-2.470) rectangle (0.000,-2.030);
\node[left,font=\footnotesize] at (-3.400,-2.250) {Cortex-A53 (i.MX8M)~\textcolor{black!55}{\scriptsize no dotprod}};
\node[right,font=\footnotesize] at (3.500,-2.250) {1.65$\times$ slower};
\end{tikzpicture}
\caption{The INT8 speedup sign is an ISA property. Bar length is the log-scaled speedup (FP32 latency $\div$ INT8 latency) for the identical ResNet-50 INT8 QDQ model on the identical ONNX Runtime CPU (MLAS) path; bars to the right of $1\times$ are faster under INT8, bars to the left are slower. The two cores with an ARM dot-product instruction gain; the two without one lose. Same data as Table~\ref{tab:c1}.}
\label{fig:c1}
\end{figure}

We ran the identical ResNet-50 INT8 QDQ model on the identical runtime (ONNX Runtime, CPU execution provider, MLAS kernels) on four CPUs, and compared against the same model in FP32 on each. Table~\ref{tab:c1} and Figure~\ref{fig:c1} report the four measurements.

\begin{table}[htbp]
\centering
\small
\begin{tabular}{L{3.8cm} L{3.0cm} L{3.0cm} L{2.6cm}}
\toprule
\textbf{CPU} & \textbf{Dot-product ISA} & \textbf{FP32 $\rightarrow$ INT8} & \textbf{INT8 effect} \\
\midrule
Cortex-A76 (Raspberry Pi 5)   & ARM \texttt{dotprod} (SDOT) & $144.95 \rightarrow 79.08$\,ms & \textbf{1.83$\times$ faster} \\
Cortex-A78AE (Jetson AGX Orin) & ARM \texttt{dotprod}        & $38.47 \rightarrow 18.22$\,ms  & \textbf{2.11$\times$ faster} \\
Core i9-10900K (x86)          & no AVX-512 VNNI             & $9.28 \rightarrow 16.34$\,ms   & \textbf{1.76$\times$ slower} \\
Cortex-A53 (i.MX8M-Nano)      & no dot-product             & $680.20 \rightarrow 1123.02$\,ms & \textbf{1.65$\times$ slower} \\
\bottomrule
\end{tabular}
\caption{The INT8 speedup sign is set by the CPU's dot-product ISA, for the identical model and runtime.}
\label{tab:c1}
\end{table}

The determinant is not the ISA \emph{family} (both ARM and x86 appear on both sides): the A53 is ARM yet slows down, the A76/A78AE are ARM and speed up. The determinant is the presence of a dot-product/accumulate instruction (ARM \texttt{dotprod}, x86 VNNI). Without it, INT8's re-quantization and widening overhead is not amortized by a faster inner product, and INT8 is a \emph{pessimization}. This has an immediate practical consequence: a fleet of heterogeneous edge CPUs cannot assume INT8 is a win; the same binary regresses on the wrong core.

% ---------------------------------------------------------------------
\section{INT8 Outputs Are Not Portable (C2) --- headline}

If two targets run the \emph{same} quantized model with the \emph{same} scales, do they produce the same predictions? For FP32 the answer is yes, exactly. For INT8 it is no.

\paragraph{FP32 is a bit-identical control.}
Across every platform pair where it was measured, FP32 top-1 predictions agree on 1000/1000 inputs. Whatever divergence we see under INT8 is therefore not a floating-point reduction-order artifact of our harness; it is specific to the integer path. We scope this bit-identity to our observed configuration (fixed thread count, a single reduction path per target): floating-point non-associativity can make FP reductions non-deterministic under different parallelization or hardware \citep{fpnonassoc2024}, so the precise claim is ``FP32 was bit-identical across these targets as measured,'' not that FP32 is portable by construction. The contrast we rely on --- FP32 identical, INT8 diverging, on the \emph{same} devices and harness --- holds regardless.

\begin{figure}[htbp]
\centering
\scriptsize
\begin{tikzpicture}[x=1cm,y=1cm]
\definecolor{cint}{RGB}{35,80,150}
\fill[black!8] (3.500,0.32) rectangle (4.375,-3.520);
\node[above,black!45,font=\scriptsize,align=center] at (3.938,0.34) {different\\integer kernel};
\draw[black!45,dashed] (8.750,0.32) -- (8.750,-3.520);
\node[above,black!45,font=\scriptsize,align=center] at (8.750,0.34) {same\\integer kernel};
\draw[black!45] (0.000,-3.520) -- (9.000,-3.520);
\draw[black!45] (1.250,-3.520) -- (1.250,-3.650);
\node[below,black!55,font=\scriptsize] at (1.250,-3.650) {940};
\draw[black!45] (3.750,-3.520) -- (3.750,-3.650);
\node[below,black!55,font=\scriptsize] at (3.750,-3.650) {960};
\draw[black!45] (6.250,-3.520) -- (6.250,-3.650);
\node[below,black!55,font=\scriptsize] at (6.250,-3.650) {980};
\draw[black!45] (8.750,-3.520) -- (8.750,-3.650);
\node[below,black!55,font=\scriptsize] at (8.750,-3.650) {1000};
\node[below,black!55,font=\scriptsize] at (5.000,-4.000) {agreeing inputs out of 1000};
\draw[black!7] (0,0.000) -- (9.000,0.000);
\node[left,font=\scriptsize,text=black] at (-0.18,0.000) {Jetson A78AE $\leftrightarrow$ Pi~5 A76};
\draw[black!65,fill=white] (8.750,0.120) circle (0.085);
\fill[cint] (8.750,-0.120) circle (0.085);
\node[right,font=\scriptsize,text=black] at (9.100,0.000) {1000};
\draw[black!7] (0,-0.620) -- (9.000,-0.620);
\node[left,font=\scriptsize,text=black] at (-0.18,-0.620) {i.MX8M A53 $\leftrightarrow$ Pi~5 A76};
\draw[black!65,fill=white] (8.750,-0.500) circle (0.085);
\fill[cint] (4.375,-0.740) circle (0.085);
\node[right,font=\scriptsize,text=black] at (9.100,-0.620) {965};
\draw[black!7] (0,-1.240) -- (9.000,-1.240);
\node[left,font=\scriptsize,text=black] at (-0.18,-1.240) {i.MX8M A53 $\leftrightarrow$ x86 i9};
\draw[black!65,fill=white] (8.750,-1.120) circle (0.085);
\fill[cint] (3.875,-1.360) circle (0.085);
\node[right,font=\scriptsize,text=black] at (9.100,-1.240) {961};
\draw[black!7] (0,-1.860) -- (9.000,-1.860);
\node[left,font=\scriptsize,text=black] at (-0.18,-1.860) {Pi~5 A76 $\leftrightarrow$ x86 i9};
\draw[black!65,fill=white] (8.750,-1.740) circle (0.085);
\fill[cint] (3.500,-1.980) circle (0.085);
\node[right,font=\scriptsize,text=black] at (9.100,-1.860) {958};
\draw[black!7] (0,-2.480) -- (9.000,-2.480);
\node[left,font=\scriptsize,text=black] at (-0.18,-2.480) {Jetson iGPU (TRT) $\leftrightarrow$ A78AE (MLAS)};
\draw[black!65,fill=white] (8.750,-2.360) circle (0.085);
\fill[cint] (3.875,-2.600) circle (0.085);
\node[right,font=\scriptsize,text=black] at (9.100,-2.480) {961};
\draw[black!7] (0,-3.100) -- (9.000,-3.100);
\node[left,font=\scriptsize,text=black!45] at (-0.18,-3.100) {DX-M1 NPU $\leftrightarrow$ host A76 $\ddagger$};
\fill[black!45] (1.125,-3.220) circle (0.085);
\node[right,font=\scriptsize,text=black!45] at (9.100,-3.100) {939};
\draw[black!65,fill=white] (0.15,-4.000) circle (0.085);
\node[right,black!55,font=\scriptsize] at (0.28,-4.000) {FP32};
\fill[cint] (1.35,-4.000) circle (0.085);
\node[right,black!55,font=\scriptsize] at (1.48,-4.000) {INT8};
\end{tikzpicture}
\caption{INT8 agreement is bimodal, and the location of the hardware boundary does not predict it. Each row is one target pair running the same model artifact: $\circ$ is FP32 top-1 agreement, $\bullet$ is INT8. FP32 is 1000/1000 wherever it was measured. INT8 is exactly 1000/1000 for the one pair that shares an integer kernel and falls in a 958--965/1000 band (shaded) for every pair that does not --- including the CPU$\leftrightarrow$accelerator pair, which sits \emph{inside} the CPU$\leftrightarrow$CPU range. The DEEPX row ($\ddagger$, gray) cannot hold the scales fixed (\S6) and is a deployment observation, excluded from the invariance claim. Same data as Table~\ref{tab:c2}.}
\label{fig:c2}
\end{figure}
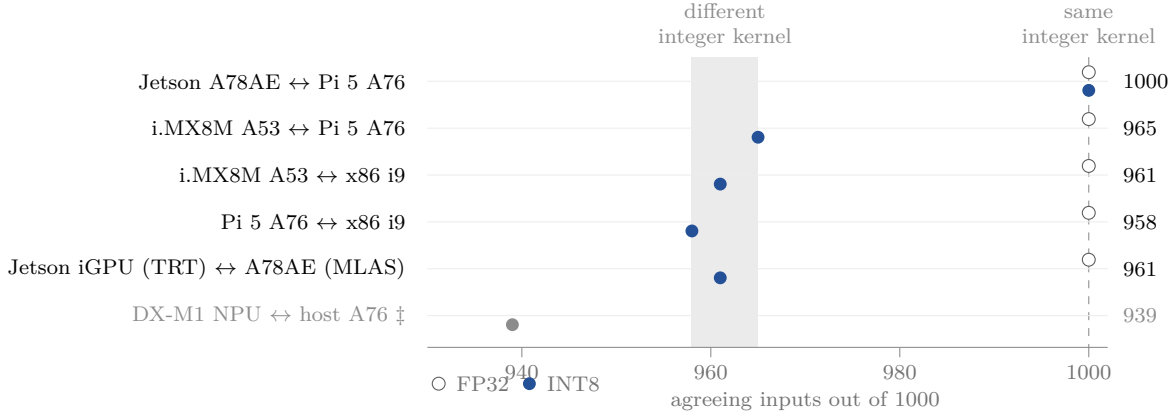

\paragraph{INT8 disagreements, ordered by kernel divergence.}
Table~\ref{tab:c2} lists every pair we measured; Figure~\ref{fig:c2} plots the same data on a single agreement axis, where the bimodality is immediate.

\begin{table}[htbp]
\centering
\small
\begin{tabular}{L{4.4cm} L{2.6cm} L{3.0cm} C{1.4cm} C{1.6cm}}
\toprule
\textbf{Comparison} & \textbf{Boundary} & \textbf{Integer kernel} & \textbf{FP32} & \textbf{INT8} \\
\midrule
Jetson A78AE vs.\ Pi~5 A76 & CPU$\leftrightarrow$CPU & \textbf{same} (MLAS SDOT) & 1000/1000 & \textbf{1000/1000} \\
i.MX8M-Nano A53 vs.\ Pi~5 A76 & CPU$\leftrightarrow$CPU & different & 1000/1000 & 965/1000 \\
i.MX8M-Nano A53 vs.\ x86 i9 & CPU$\leftrightarrow$CPU & different & 1000/1000 & 961/1000 \\
Raspberry Pi~5 A76 vs.\ x86 i9 & CPU$\leftrightarrow$CPU & different & 1000/1000 & 958/1000 \\
Jetson iGPU (TensorRT) vs.\ A78AE (MLAS) & \textbf{CPU$\leftrightarrow$accel.} & different & 1000/1000 & \textbf{961/1000} \\
DEEPX DX-M1 (NPU) vs.\ host A76 CPU~$\ddagger$ & \textbf{CPU$\leftrightarrow$vendor-NPU} & different, \emph{and different scales} & --- & 939/1000 \\
\bottomrule
\end{tabular}
\caption{INT8 top-1 agreement between targets running the same model artifact. Agreement is bimodal: 1000/1000 when the integer kernel is shared, 958--965/1000 when it is not, independent of where the hardware boundary falls. FP32 is 1000/1000 for every pair measured. $\ddagger$~excluded from the invariance claim (scales not held fixed; see text).}
\label{tab:c2}
\end{table}

\textbf{The result is an invariance, not a gradient.} Across the five pairs that share one model artifact and its embedded scales, agreement takes essentially two values: \textbf{1000/1000 when the two targets execute the same integer kernel, and 958--965/1000 when they do not.} Nothing else predicts it --- in particular, the \emph{location} of the hardware boundary does not. The CPU$\leftrightarrow$accelerator pair (961/1000) falls inside the range spanned by the three CPU$\leftrightarrow$CPU pairs (958--965/1000), and the one pair that agrees perfectly is itself a CPU$\leftrightarrow$CPU pair spanning two different SoCs, two different boards, and two ONNX Runtime versions (1.23.2 and 1.28.0). Sharing the integer kernel is \emph{sufficient} for identical predictions; crossing a chip boundary is \emph{not sufficient} to break them. What breaks them is a different re-quantization epilogue.

The CPU$\leftrightarrow$GPU row makes the mechanism explicit: it uses the \emph{same} \texttt{resnet50\_int8\_qdq.onnx} to build the TensorRT engine that the CPU EP runs, so the QDQ scales are bit-identical and the only free variable is the integer kernel (TensorRT's INT8 kernels vs.\ MLAS's) --- yet the two disagree on 39/1000 inputs. The divergence is kernel-specific rounding and accumulation, not hardware noise or a scale mismatch.

\textbf{The vendor-NPU row is excluded from the invariance claim.} It is the one comparison here that cannot hold scales fixed: the DEEPX compiler rejects an external QDQ graph outright (\S6) and runs its own PTQ, so the pair differs simultaneously in integer kernel, quantizer, and calibration set. Per the controlled-comparison principle of \S3.3 we therefore report 939/1000 as a \emph{deployment} observation --- what a fielded CPU-plus-vendor-NPU pairing actually delivers --- and not as a kernel comparison. It is consistent with, but does not evidence, the invariance above.

\paragraph{Accuracy hides it.}
The disagreements are net-neutral: on the Jetson iGPU, accuracy-valid INT8 top-1 is 0.7620 --- \emph{lossless} versus its own FP32 and higher than the CPU MLAS INT8 top-1 of 0.7500 --- even though the two INT8 paths flip predictions on dozens of individual inputs. A standard ``accuracy after quantization'' report would show no problem. The portability failure is only visible when you compare \emph{predictions per input across targets}.

\paragraph{Relation to concurrent work.}
The mechanism behind these disagreements --- divergence introduced in the re-quantization \emph{epilogue} (scale application and output rounding) after an exact INT32 accumulate --- was localized concurrently by Chen \citep{chen2026integeralibi, chen2026deterministic}, who swapped INT8 GEMM kernels (CUTLASS vs.\ Triton) \emph{on a single GPU} for LLMs and showed that power-of-two scales restore bit-identical cross-kernel agreement. We do not claim to discover this mechanism, and we cite that work as concurrent prior art. Our result is complementary and, for deployment, more consequential in three ways: the divergence persists across \emph{different physical devices} (a dot-product CPU vs.\ a non-dot-product CPU, a CPU vs.\ a GPU built from the same scales, a CPU vs.\ a third-party vendor NPU); it holds for vision and detection models rather than LLMs; and it is measured against an FP32 bit-identical control on the same devices. Where Chen asks whether two kernels on one GPU agree, we ask whether a fielded fleet of heterogeneous targets agrees --- the question a cross-module automotive platform actually poses.

\paragraph{We tested that mitigation across a physical boundary; it does not transfer.}
Chen's power-of-two result is demonstrated for two kernels on one GPU. We forced \emph{every} Q/DQ scale in \texttt{resnet50\_int8\_qdq.onnx} to a power of two --- making the requant multiplier $M = (s_a \cdot s_w)/s_\mathrm{out}$ a pure shift $2^{k_a + k_w - k_\mathrm{out}}$ that any kernel \emph{should} implement identically --- rebuilt the model (weights re-quantized to the new scales; the zero-points are already all zero, so $M$ is the only divergence term), and re-ran the x86 $\leftrightarrow$ Pi~5 (A76) pair. It fails, and informatively: agreement does not rise to 1000/1000 but \emph{drops below the baseline}, 958/1000 $\to$ \textbf{869/1000} (ceil rounding, scales raised by ${\le}1$ bit) and \textbf{919/1000} (nearest, ${\le}\sqrt{2}$) --- a gate FAIL on both arms. MLAS does not special-case $M = 2^k$ into a shared exact shift; forcing powers of two instead coarsens the requantization grid and \emph{enlarges} the epilogue divergence it was meant to remove --- the per-element cross-kernel $|\Delta\mathrm{logit}|$ grows $\times2.35$ (nearest) to $\times4.34$ (ceil), monotonically with the flip count. A mechanism probe (report \texttt{stage5\_pot\_scales\_report.html}) separates the two possible causes: the divergence-magnitude increase is confirmed and amplified, while the competing tie-breaking hypothesis --- that power-of-two scales create more exact \texttt{.5} ties for half-even vs.\ half-away kernels to split --- is \emph{refuted}, the near-tie population staying essentially flat (41 $\to$ 36 / 45 across baseline / ceil / nearest) and moving \emph{opposite} to the flip count. All figures are exact rather than point estimates: 30 repeat runs are bit-identical. Chen's single-GPU finding is not contradicted; our measurement shows only that it does not carry across the x86$\leftrightarrow$A76 boundary --- the power-of-two mitigation whose scope \S11 bounds.

\paragraph{Why it matters.}
A quantized model is often treated as a deterministic function $f(x)$. C2 says that once you change the target, $f$ changes on a measurable fraction of inputs, silently, with no accuracy signal. For automotive systems this is the crux: dual-compute redundancy and cross-module consistency checks assume two units computing the same input agree; C2 shows that assumption holds under FP32 but not under INT8 across heterogeneous integer kernels. The invariance sharpens what the design constraint is: determinism does not degrade gradually with hardware distance, so it cannot be bought back by choosing a ``closer'' second target. It is binary in the integer kernel. Two units agree per-input only if they run the same integer kernel --- which, across a heterogeneous redundant architecture, is precisely the property the architecture was chosen to \emph{avoid}.

% ---------------------------------------------------------------------
\section{Vendors Own Quantization: Two Failure Modes (C3)}

C1--C2 assume you can even \emph{run} your quantized model on the target. On vendor NPUs, you frequently cannot --- the accelerator insists on quantizing the model itself.

\paragraph{Qualcomm Hexagon HTP --- silent.}
Submitting an externally quantized ONNX-QDQ ResNet-50 to Qualcomm AI Hub compiles, profiles, and runs on-device with no error, but the HTP ignores the external QDQ scales and on-device top-1 collapses from 0.75 to \textbf{0.005}. The same ONNX runs correctly on x86 CPU (0.753), and the FP32/fp16 path on HTP is faithful (0.745) --- so the failure is specific to \emph{externally supplied} INT8 scales being discarded. The correct path is the vendor's own \texttt{submit\_quantize\_job} (HTP-native QDQ), which recovers top-1 to \textbf{0.735} and is \emph{faster and leaner} than the external-QDQ engine (748\,$\mu$s vs.\ 1052\,$\mu$s).

\paragraph{DEEPX DX-M1 --- loud.}
Feeding the same class of externally quantized graph to the DEEPX compiler produces a hard error --- \texttt{GraphStructureError: 106 isolated node(s)} $\rightarrow$ \texttt{InternalError}, with no engine emitted. The native path (supply FP32; let the DEEPX compiler run its own PTQ) compiles cleanly and reaches top-1 \textbf{0.7660}, lossless-grade.

\paragraph{One root cause, opposite symptoms.}
Qualcomm fails \emph{open} (runs a silently broken model --- dangerous, because a broken model can ship) and DEEPX fails \emph{closed} (refuses to build --- safe, because nothing broken can ship). Both encode the same rule: the accelerator, not your toolchain, owns the quantization. The deployment consequence is that a quantization you validated on one target is not a portable artifact to a vendor NPU at all --- reinforcing C2 from the deployability side, and adding a concrete safety-relevant hazard in the Qualcomm case (a silently wrong INT8 model that passes compile and profile).

% ---------------------------------------------------------------------
\section{Bottleneck Regimes Are Set by Output Size, Not Hardware (C4)}

Reasoning about ``is this NPU fast enough'' usually starts from compute (FLOPs/TOPS). On a PCIe-attached edge NPU we find the latency \emph{regime} --- and whether adding cores helps at all --- is set by the model's output (device-to-host, D2H) transfer size, on the same device and runtime. Table~\ref{tab:c4} puts three models on that one accelerator.

\begin{table}[htbp]
\centering
\small
\begin{tabular}{L{3.4cm} L{2.0cm} L{5.2cm} L{3.4cm}}
\toprule
\textbf{Model (on DX-M1)} & \textbf{Output} & \textbf{Regime} & \textbf{Multi-core scaling} \\
\midrule
ResNet-50 & 4\,KB           & \textbf{compute-bound} (compute 2.77\,ms $\gg$ D2H 0.11\,ms) & $2.19\times$ near-linear on 3 cores \\
YOLO26n   & 2.82\,MB (raw head) & \textbf{D2H-bound} (D2H 21.81\,ms $\gg$ compute 9.0\,ms) & $1.00\times$ flat \\
YOLOv5s   & 5.48\,MB        & \textbf{D2H-bound}, worse & $1.02\times$ flat; lightest compute (2.59\,ms) yet \textbf{26$\times$ slower than ResNet-50} \\
\bottomrule
\end{tabular}
\caption{On one PCIe-attached edge NPU, the latency regime and multi-core scaling track output/D2H size, not compute.}
\label{tab:c4}
\end{table}

The YOLOv5s row isolates the causal variable: it has the \emph{smallest} compute of the three yet is by far the slowest --- 41.01\,fps on three cores against ResNet-50's 1078.93\,fps, a $26.3\times$ gap in the wrong direction --- because it has the largest output to move across the PCIe Gen2$\times$1 link. Compute does not predict the regime; output/D2H size does.

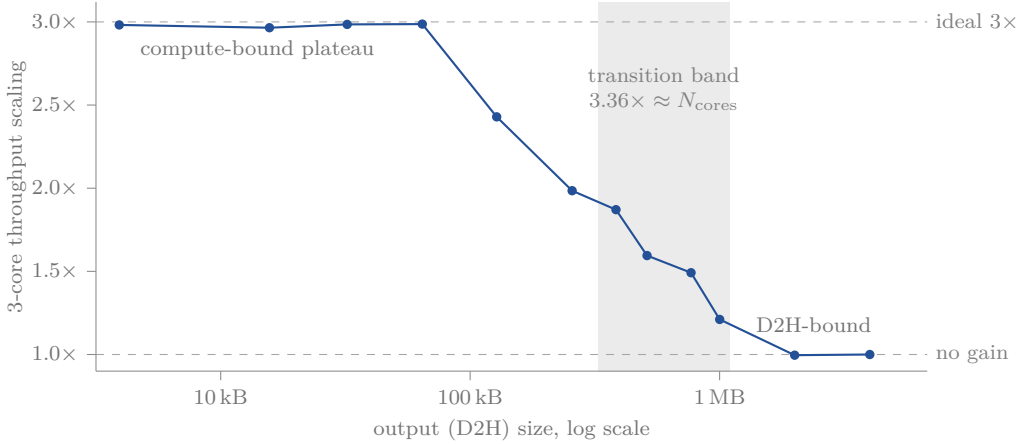
\begin{figure}[htbp]
\centering
\footnotesize
\begin{tikzpicture}[x=1cm,y=1cm]
\definecolor{cint}{RGB}{35,80,150}
\fill[black!8] (6.645,0) rectangle (8.381,4.884);
\node[black!50,font=\scriptsize,align=center,above] at (7.513,3.344) {transition band\\$3.36\times \approx N_{\mathrm{cores}}$};
\draw[black!35,dashed] (0,4.620) -- (10.997,4.620) node[right,black!50,font=\scriptsize] {ideal $3\times$};
\draw[black!35,dashed] (0,0.220) -- (10.997,0.220) node[right,black!50,font=\scriptsize] {no gain};
\draw[black!45] (0,0) -- (0,4.884);
\draw[black!45] (0,0.220) -- (-0.12,0.220);
\node[left,black!55,font=\scriptsize] at (-0.12,0.220) {1.0$\times$};
\draw[black!45] (0,1.320) -- (-0.12,1.320);
\node[left,black!55,font=\scriptsize] at (-0.12,1.320) {1.5$\times$};
\draw[black!45] (0,2.420) -- (-0.12,2.420);
\node[left,black!55,font=\scriptsize] at (-0.12,2.420) {2.0$\times$};
\draw[black!45] (0,3.520) -- (-0.12,3.520);
\node[left,black!55,font=\scriptsize] at (-0.12,3.520) {2.5$\times$};
\draw[black!45] (0,4.620) -- (-0.12,4.620);
\node[left,black!55,font=\scriptsize] at (-0.12,4.620) {3.0$\times$};
\node[rotate=90,black!55,font=\scriptsize] at (-1.05,2.442) {3-core throughput scaling};
\draw[black!45] (0,0) -- (10.997,0);
\draw[black!45] (1.650,0) -- (1.650,-0.12);
\node[below,black!55,font=\scriptsize] at (1.650,-0.12) {10\,kB};
\draw[black!45] (4.950,0) -- (4.950,-0.12);
\node[below,black!55,font=\scriptsize] at (4.950,-0.12) {100\,kB};
\draw[black!45] (8.250,0) -- (8.250,-0.12);
\node[below,black!55,font=\scriptsize] at (8.250,-0.12) {1\,MB};
\node[below,black!55,font=\scriptsize] at (5.499,-0.52) {output (D2H) size, log scale};
\draw[cint,thick] (0.308,4.580) -- (2.295,4.543) -- (3.323,4.587) -- (4.317,4.591) -- (5.301,3.364) -- (6.299,2.387) -- (6.879,2.136) -- (7.290,1.529) -- (7.872,1.302) -- (8.251,0.684) -- (9.243,0.211) -- (10.237,0.220);
\fill[cint] (0.308,4.580) circle (0.062);
\fill[cint] (2.295,4.543) circle (0.062);
\fill[cint] (3.323,4.587) circle (0.062);
\fill[cint] (4.317,4.591) circle (0.062);
\fill[cint] (5.301,3.364) circle (0.062);
\fill[cint] (6.299,2.387) circle (0.062);
\fill[cint] (6.879,2.136) circle (0.062);
\fill[cint] (7.290,1.529) circle (0.062);
\fill[cint] (7.872,1.302) circle (0.062);
\fill[cint] (8.251,0.684) circle (0.062);
\fill[cint] (9.243,0.211) circle (0.062);
\fill[cint] (10.237,0.220) circle (0.062);
\node[black!55,font=\scriptsize,below right] at (0.473,4.488) {compute-bound plateau};
\node[black!55,font=\scriptsize,above left] at (10.373,0.396) {D2H-bound};
\end{tikzpicture}
\caption{The compute-bound $\to$ D2H-bound transition with NPU compute held fixed. All 12 synthetic sweep points on the DEEPX DX-M1: 3-core throughput scaling against output (D2H) size, at a constant 2.70\,ms 1-core inference (spread 4.9\%). Scaling holds a $2.97$--$2.99\times$ plateau while the output stays below ${\approx}63$\,KB, then descends to $1.00\times$. The shaded band runs from the link-saturation onset (326{,}377\,B, where D2H $=$ compute$/N$) to the single-inference crossover (1{,}095{,}949\,B, where D2H $=$ compute); its $3.36\times$ width equals the core count $N=3$, because $N$ cores share one D2H link. Five of these points are listed numerically in Table~\ref{tab:c4sweep}.}
\label{fig:c4}
\end{figure}

\textbf{A controlled single-variable sweep traces the transition curve.} The three models above bracket the two regimes but cannot trace the boundary between them, because they differ in compute as well as output. We remove that residual confound with a synthetic sweep on the same device and runtime: a fixed heavy convolutional trunk holds NPU compute constant (1-core inference p50 $=2.70$\,ms, spread 4.9\% across the sweep) while a 4-channel bottleneck feeding a $1\times1$ ``expander'' head varies \emph{only} the output tensor, over $1020\times$ (3{,}920\,B $\to$ 3{,}999{,}968\,B; 12 points). Holding compute fixed and sweeping output alone, 3-core throughput scaling sits on a flat compute-bound plateau of $2.97$--$2.99\times$ while the output stays below ${\approx}63$\,KB, then descends monotonically to $1.00\times$ (D2H-bound), and the per-core job distribution migrates 33/33/33 $\to$ 98/2/0 as the one shared PCIe link progressively starves all but one core (reproducing the YOLO26n 472/28/2 signature). The single-inference crossover --- where D2H equals the fixed 2.70\,ms compute --- falls at ${\approx}1.05$\,MB of output. Table~\ref{tab:c4sweep} lists five representative points of the sweep; Figure~\ref{fig:c4} plots all twelve.

\begin{table}[htbp]
\centering
\small
\begin{tabular}{R{3.6cm} R{2.0cm} L{4.6cm}}
\toprule
\textbf{Output bytes (fixed 2.70\,ms compute)} & \textbf{D2H p50} & \textbf{3-core scaling} \\
\midrule
3{,}920     & 0.141\,ms & $2.98\times$ (compute-bound plateau) \\
256{,}368   & 0.776\,ms & $1.99\times$ (transition band) \\
511{,}952   & 1.367\,ms & $1.59\times$ (transition band) \\
1{,}000{,}384 & 2.429\,ms & $1.21\times$ (transition band) \\
1{,}999{,}984 & 8.892\,ms & $1.00\times$ (D2H-bound) \\
\bottomrule
\end{tabular}
\caption{Controlled sweep: with compute held fixed, 3-core scaling descends past the compute-bound plateau as output/D2H size grows, tracing the compute-bound $\to$ D2H-bound transition. Output sizes are exact byte counts, not rounded KB/MB, because the two regime thresholds fall between conventional round numbers.}
\label{tab:c4sweep}
\end{table}

\textbf{The regime boundary is a band whose width equals the core count.} The transition is not a knife-edge but a band bounded by two thresholds: multi-core scaling begins to break when D2H reaches compute$/N$ (${\approx}319$\,KB --- the one shared link can no longer feed all $N$ cores), and a \emph{single} inference turns D2H-bound when D2H reaches compute (${\approx}1.05$\,MB). Their ratio is $N$: we measure a $3.36\times$ band on a 3-core device, because $N$ cores share one D2H link, so aggregate throughput saturates at $1/N$ of the per-core D2H that bounds one inference. This gives a closed-form provisioning rule --- from compute time and link bandwidth alone, an $N$-core accelerator behind a bus stops scaling once the output exceeds ${\approx}(\text{compute-time} \times \text{link-bandwidth})/N$ --- turning the qualitative ``provision by data movement'' into a quantitative threshold. (Thresholds are linear-fit extrapolations, 2.456\,ms/MB, specific to this Gen2$\times$1 link; the band-width $=N$ ratio is bandwidth-independent. See \S10.)

We observe a third regime with the transformer detector DETR, where the DEEPX compiler auto-splits the graph and leaves the transformer on the host CPU in FP32: end-to-end 1036.34\,ms decomposes as host-CPU transformer FP32 910.6\,ms (87.9\%) $\gg$ D2H 57.28\,ms $\gg$ NPU INT8 41.11\,ms (4.0\%) $\gg$ H2D 6.97\,ms --- \textbf{host-CPU-compute-bound}. (The four stages sum to 1015.96\,ms, which matches the runtime's own end-to-end p50 of 1017.53\,ms; the 1036.34\,ms figure is the benchmark harness's wall-clock, so the ${\approx}20$\,ms excess is harness overhead rather than unaccounted device time.) Three models on one accelerator thus exhibit three different bottlenecks (NPU-compute, PCIe-D2H, host-CPU-compute). For context, in its favorable (compute-bound) regime the same NPU delivers large wins over the host CPU --- e.g., YOLO26n throughput 91.51\,fps vs.\ 8.01\,fps on the A76 ($\times11.42$) and host-side perf/watt $\times29.29$ --- but those wins evaporate the moment the model's output pushes it into the D2H-bound regime. The design rule: for edge accelerators behind a bus, provision and partition by data movement, not by TOPS.

% ---------------------------------------------------------------------
\section{When INT8 Breaks Down: Transformers and the Granularity Lever (supporting)}

C1--C3 use CNNs, where INT8 is (kernel permitting) nearly lossless. Transformers are the stress test and reinforce C2's thesis that \emph{activations}, not ops, are where INT8 portability breaks.

\subsection{The collapse, and what does \emph{not} explain it}

DETR INT8 collapses hard and reproducibly: FP32 mAP $0.4207 \rightarrow$ INT8 $0.2402$ ($-42.9$\%) on a discrete GPU (ORT \texttt{quantize\_static}, QDQ format, per-channel \texttt{QInt8} weights and \textbf{per-tensor} \texttt{QInt8} activations, MinMax calibration over 100 images, CUDA EP, COCO val2017 5{,}000 images), cross-confirmed on Jetson with symmetric re-quantization ($0.4237 \rightarrow 0.2383$, $-43.8$\%), with small-object mAP down $77$--$85$\%.

\textbf{Op selection is not the lever.} Leaving all 36 attention-score matmuls in FP barely moves the result ($0.2402 \rightarrow 0.2438$, $\mathbf{+0.0036}$~mAP). Three of the four exclusion patterns routinely recommended for transformer PTQ are in fact no-ops on this graph: DETR contains no GELU, and ORT's QDQ quantizer does not touch Softmax or LayerNorm to begin with. Only the attention-matmul exclusion is even applicable, and it recovers 2\% of the gap.

\textbf{Nor is the damage owned by one half of the network.} Quantizing each half alone shows both halves collapse on their own (Table~\ref{tab:ablation}):

\begin{table}[htbp]
\centering
\small
\begin{tabular}{L{6.6cm} C{2.2cm} C{2.6cm}}
\toprule
\textbf{Configuration (ORT, COCO val2017 5{,}000)} & \textbf{mAP} & \textbf{$\Delta$ vs.\ FP32} \\
\midrule
FP32                                    & 0.4207 & --- \\
Backbone only INT8 (53 Convs)           & 0.2653 & $\mathbf{-36.9}$\% \\
Transformer only INT8 (137 nodes)       & 0.2391 & $-43.2$\% \\
All INT8 (190 nodes)                    & 0.2402 & $-42.9$\% \\
\bottomrule
\end{tabular}
\caption{Two-way ablation: each half of DETR collapses on its own, and the two losses are strongly sub-additive --- the damage is distributed, not localized.}
\label{tab:ablation}
\end{table}

The two halves' losses are strongly \textbf{sub-additive} --- either half alone reproduces most of the full collapse, and their separate losses sum to far more than the whole. The damage is therefore distributed across the network rather than localized to a fragile subgraph. What the two halves share is not an op type but the quantizer's \emph{activation} treatment: one per-tensor MinMax scale per activation tensor. That is the axis \S8.3 isolates.

The activation-granularity lever (SmoothQuant) recovers 59.9\% of the gap in a torch fake-quant setting but only ${\sim}9$\% on-device --- because the only on-device-buildable INT8 path quantizes Gemms only (attention/LayerNorm/Softmax stay FP16), so the lever cannot reach the dominant residual.

\subsection{A toolchain that declines: no collapse by avoidance}

The DEEPX compiler produces \emph{no} transformer INT8 collapse on the same model (mAP $0.4377 \rightarrow 0.4385$, $+0.0008$, within noise). The reason is not better transformer quantization: the compiler auto-partitions the 708-node graph and assigns only the CNN backbone and the first encoder self-attention block to the NPU in INT8, leaving the remaining encoder layers, all six decoder layers, the FFNs and the heads on the \textbf{host CPU in FP32}. The handoff tensor is the giveaway --- 23.99\,MB of FP32 attention state crosses back to the host per frame.

``No collapse'' and ``collapse'' are thus two sides of one fact: transformer activations do not survive INT8, so a toolchain either refuses (no loss, no speedup --- \S7 shows the resulting pipeline is host-CPU-bound at 1.04\,s end-to-end) or forces it (speedup, large loss). This is the portability thesis again, in the accuracy dimension.

\subsection{The quantizer does not port either}

\S8.2 leaves a matched pair we can exploit. Both toolchains were, in effect, asked to run the \emph{same nominal recipe} --- INT8 the CNN backbone, keep the transformer in FP32 --- and both report an FP32 reference computed with the transformer in FP32. Their answers differ by 37.1 points of relative mAP (Table~\ref{tab:quantizer}).

\begin{table}[htbp]
\centering
\small
\begin{tabular}{L{3.3cm} L{4.6cm} L{5.2cm}}
\toprule
 & \textbf{ORT (\S8.1, discrete GPU)} & \textbf{DEEPX \texttt{dx\_com} 2.4.0 (DX-M1)} \\
\midrule
INT8 scope        & backbone Conv $\times 53$ (\texttt{input\_projection} excluded) & backbone \textbf{+ \texttt{input\_projection} + first encoder self-attn through its Softmax} \\
Weights           & per-channel \texttt{QInt8} & \texttt{wbit8} \\
\textbf{Activations} & \textbf{per-tensor \texttt{QInt8}, MinMax, 100 images} & \textbf{\texttt{abit8}, EMA, 100 in-domain COCO images} \\
Calib.\ input shape & dynamic axes (varies per image) & fixed $800\times1066$, bit-identical \texttt{.npy} \\
Evaluation        & COCO val2017, 5{,}000 images & COCO val2017, first 500 images \\
FP32 $\rightarrow$ INT8 & $0.4207 \rightarrow 0.2653$ & $0.4377 \rightarrow 0.4385$ \\
\textbf{Relative $\Delta$} & $\mathbf{-36.9}$\% & $\mathbf{+0.2}$\% (within noise) \\
\bottomrule
\end{tabular}
\caption{The same nominal recipe (``INT8 the backbone, keep the transformer FP32'') under two vendors' quantizers. The INT8 scope runs \emph{against} the outcome --- DEEPX quantizes strictly more and loses strictly less --- leaving the activation calibrator as the operative difference.}
\label{tab:quantizer}
\end{table}

The obvious explanations do not survive contact with the rows. The INT8 \emph{scope} runs the wrong way: DEEPX quantizes a strictly larger portion of the graph --- including the input projection and an entire attention block that ORT leaves in floating point --- and loses less. Evaluation subset (5{,}000 vs.\ 500) and fixed-vs-dynamic resolution shift a mAP by ones of percent, not by 37.1 points.

What remains is the activation calibrator. A per-tensor MinMax scale is set by the single largest activation observed during calibration, so one outlier image dilates the range and quantizes every other activation coarsely; an EMA calibrator damps exactly that. The ORT path amplifies the effect further by calibrating over dynamic input shapes, so the 100 calibration images do not even share a common activation geometry. This is the \S8.1 lever --- activation granularity and variance, not op selection --- now observed \emph{across two vendors' quantizers} rather than within one.

The deployment consequence is a fourth portability axis, and in practice the sharpest. C2 (\S5) shows that holding the scales fixed still leaves the \emph{outputs} target-dependent. \S8.3 shows that when you \emph{cannot} hold the scales fixed --- which \S6 establishes is the normal case on a vendor NPU --- even the \emph{accuracy} of a nominally identical recipe becomes a property of the vendor's calibrator rather than of your model. ``We quantized the backbone and it was fine'' is not a transferable statement.

\emph{(Caveat. The two columns are not comparable in absolute mAP: different evaluation subsets, resolutions, and runtimes. Only the within-column FP32$\rightarrow$INT8 relative deltas are compared --- which is precisely the quantity that differs by 37.1 points, and each column's FP32 reference is measured on its own pipeline.)}

\emph{(Accelerator note. On the Jetson NVDLA v2, INT8 is not merely preferred but mandatory: DLA is an INT8-only datapath (FP16 is $13.87\times$ slower) and is the perf-per-watt leader (51.29\,inf/s/W, ${\sim}1.55\times$ the iGPU at roughly half the power) for CNNs, but fragments on transformers --- the identical \texttt{-{}-useDLACore=0 -{}-allowGPUFallback} recipe that gives ResNet-50 a clean 2-fallback offload gives DETR 326 DLA layers, 404 GPU-fallback layers and 16 ForeignNodes, at 398.64\,ms versus 13.28\,ms for the same model in FP16 on the iGPU ($\mathbf{30\times}$ slower, FP16 vs.\ FP16). The accelerator's ``preferred precision'' is itself a non-portable, model-dependent property.)}

% ---------------------------------------------------------------------
\section{Methodology Pitfalls and Silent Failures (C8)}

The measurements above were only trustworthy after we removed a series of silent errors that a normal pipeline would not surface. We report them because they bound what any single-number quantization result means.

\begin{itemize}
\item \textbf{Subset evaluation inflates accuracy.} A 1000-image ImageNet subset overstated top-1 by \textbf{+9.77 percentage points on average} versus the full 50k val set --- a mean over eight model configurations, spanning $+8.41$ to $+10.39$\,pp, so the inflation cannot be corrected away by a constant offset without reordering the models. It also flipped the sign of three quantization deltas and 5 of 13 significance verdicts. Accuracy claims here are either on full val or explicitly flagged as relative-on-subset.
\item \textbf{Preprocessing dominates the quantization delta.} The choice of resize/crop (squash vs.\ torchvision) changed top-1 by \textbf{$-$1.07\,pp}, roughly \textbf{9$\times$} the $-0.12$\,pp cost of the quantization itself. A quantization number is meaningless without a fixed, reported preprocessing.
\item \textbf{SQNR does not predict accuracy.} Per-layer SQNR had essentially no rank correlation with the top-1 delta (Spearman $\rho = -0.030$ over 21 layers, 50k val); the common practice of ranking layers by SQNR to guide mixed precision is not supported here.
\item \textbf{Silent fallbacks are everywhere.} External QDQ silently falling back to CPU/FP32; a \texttt{TensorrtExecutionProvider} that is listed as available yet quietly runs on the CPU because \texttt{libnvinfer.so.10} was off the loader path (p50 11.83\,ms, i.e.\ CPU-class, versus 0.41\,ms once fixed); a zero-copy output buffer aliasing bug that collapsed top-1 to 0.0014 with no error; an opset down-convert that ``succeeds'' (exit 0) while producing an invalid graph; the pip TensorRT wheel shipping without the \texttt{trtexec} binary the tutorials assume. Each of these produces a plausible-looking number that is wrong.
\end{itemize}

% ---------------------------------------------------------------------
\section{Threats to Validity}

We state the study's limits plainly; several are properties of a measurement-first project and bound our claims to \emph{relative} comparisons.

\begin{itemize}
\item \textbf{Single-run latencies without confidence intervals.} Most latencies are p50 or single-run. We do not claim differences smaller than a few percent; the headline results (sign flips of $1.6$--$2.1\times$, regime differences of ${>}20\times$) are far outside plausible run-to-run noise, but the smaller ones should be read as directional.
\item \textbf{Subset accuracy.} Several accuracy and agreement numbers are on subsets (200--1000 images). This study itself quantifies the risk (\S9, $+9.77$\,pp mean inflation); the agreement counts (958--1000/1000) are on fixed 1000-image bundles and are internally comparable, but not comparable to full-val absolute accuracy.
\item \textbf{$n=1$ per hardware class.} One unit per class. We claim ``this representative device,'' never ``all A76'' or ``all x86.''
\item \textbf{Version confounds, and the controls that bound them.} Runtime versions differ across targets and cannot be equalized: the four CPUs ran ONNX Runtime 1.17.1 (A53), 1.23.2 (A78AE and x86 i9) and 1.28.0 (Pi~5), so \emph{every} cross-platform pair in C1 and C2 is also a cross-machine comparison, and three of the four CPU$\leftrightarrow$CPU pairs are cross-version as well. Two controls bound what the version difference can explain. (i)~FP32 predictions are 1000/1000 across all three ORT versions, so version alone does not perturb predictions on this graph. (ii)~The one INT8 pair that agrees perfectly (A78AE$\leftrightarrow$Pi~5, 1000/1000) is itself cross-version (1.23.2 vs.\ 1.28.0), while a pair that disagrees (A53$\leftrightarrow$x86, 961/1000) is cross-version too --- version does not separate the two outcomes, the integer kernel does. For C1, the sign flip survives within a single version: the A78AE and the x86 i9 both ran 1.23.2, and INT8 made one $2.11\times$ faster and the other $1.76\times$ slower. We nonetheless report the version of each target and treat any residual version effect as a limit on absolute latencies, not on the sign or the agreement invariance.
\item \textbf{Relative, not absolute.} Batch size, input resolution, and evaluation subset differ across sections; absolute latency/accuracy are not cross-comparable. All claims are within-comparison relative deltas.
\item \textbf{Power-measurement gap.} Some perf-per-watt figures use a host-side power boundary because on-board/M.2 card power (upstream of the accessible rail) or DLA power (not captured by the GPU utilization counter) could not be isolated; we report the measurement boundary alongside each figure.
\item \textbf{Init-weight models excluded from accuracy.} The BEV capstone models ran with initialization weights (public weights unavailable), so their mAP is ${\sim}0$ by construction and is used only for latency/engine-size characterization, never for accuracy claims. The \S7 regime-transition sweep likewise uses synthetic random-weight models with a synthesized input; they are valid only for latency/regime, never accuracy.
\item \textbf{Extrapolated, link-specific thresholds (\S7).} The two transition thresholds (${\approx}319$\,KB, ${\approx}1.05$\,MB) are extrapolations of a linear D2H fit (2.456\,ms/MB), not directly measured points, and their \emph{absolute} positions are specific to this host's PCIe Gen2$\times$1 link (a wider link shifts them). The structural result --- a transition band of width $=$ core count --- is link-bandwidth-independent, since it follows only from $N$ cores sharing one D2H link.
\item \textbf{Vendor scope.} Vendor-NPU findings cover Qualcomm and DEEPX; other automotive NPUs (TI, Renesas) were not available and are left to future work.
\item \textbf{Power-of-two mitigation --- scope of the negative result (\S5).} The finding that forcing power-of-two scales does not restore cross-device bit-identity --- and in fact makes agreement worse --- is measured on one model (\texttt{resnet50\_int8\_qdq.onnx}), two rounding modes (ceil, nearest), and one physical boundary (x86 no-VNNI $\leftrightarrow$ A76 SDOT, both MLAS). It is an input/output measurement, not kernel introspection: the mechanism attribution (grid coarsening enlarges the epilogue divergence; tie-breaking does not) rests on the mutual consistency of the $|\Delta\mathrm{logit}|$ magnitudes, the weight-saturation counts (ceil 0 vs.\ nearest 41{,}447), and the decision-margin distributions across the two rounding modes, not on reading the kernel source. It refutes the \emph{transfer} of Chen's single-GPU result to this boundary; it does not refute that result in its own setting, and a different boundary (CPU$\leftrightarrow$accelerator, CPU$\leftrightarrow$vendor-NPU) or a kernel pair that \emph{does} implement $M=2^k$ as an exact shift could behave differently.
\end{itemize}

% ---------------------------------------------------------------------
\section{Conclusion}

Across seven hardware classes we find that INT8 quantization is not portable on any of the three axes the operational folklore assumes. Its \emph{speedup} can be negative and its sign is set by the CPU's dot-product ISA. Its \emph{numerics} are not portable: identical scales produce disagreeing predictions across CPU$\leftrightarrow$CPU and CPU$\leftrightarrow$accelerator boundaries, while FP32 stays bit-identical --- an accuracy-invisible loss of determinism. Its \emph{deployability} is gated by the target vendor, which owns quantization and rejects a bring-your-own QDQ graph either silently (dangerously) or loudly. We add that edge-NPU latency regimes are governed by data movement, not compute.

The practical recommendations are concrete: re-validate INT8 per target rather than once; treat vendor-native quantization as mandatory, not optional; provision accelerators by output/data-movement size; and, for safety-relevant or redundant automotive compute, do not assume two heterogeneous units running the same INT8 model agree per input --- they do under FP32 and may not under INT8. Where per-input cross-target determinism is required, constraining quantization to power-of-two scales --- shown to restore bit-identical cross-kernel agreement in the single-GPU LLM setting \citep{chen2026deterministic} --- is an appealing mitigation, but we tested it across a \emph{physical} device boundary and it did not transfer: forcing every scale to a power of two on our INT8 ResNet-50 left the x86$\leftrightarrow$A76 pair \emph{further} from agreement than the unmodified baseline (958/1000 $\to$ 869/1000 for ceil rounding, 919/1000 for nearest), because these two MLAS kernels do not implement $M = 2^k$ as a shared exact shift and the coarsened scale grid enlarges the epilogue divergence rather than removing it (\S5). Whether it transfers across other boundaries (CPU$\leftrightarrow$accelerator, CPU$\leftrightarrow$vendor-NPU), or with a kernel pair that \emph{does} share an exact power-of-two shift, remains open; on the evidence here, per-target re-validation --- not a scale constraint --- is the dependable path. These findings also motivate our follow-on work characterizing a heterogeneous multi-module automotive compute platform, where the inter-module data-movement bottleneck (a level up from \S7) and cross-module INT8 consistency (a level up from \S5) become first-order system design constraints.

\paragraph{Artifact availability.} Measurement scripts and 32 HTML reports are released with this paper; Appendix~\ref{app:map} maps every numbered claim to the report and script that produced it. Available at \url{https://github.com/yyshin-katech/embedded-ai-quantization-guide/tree/paper1-v1}.

% ---------------------------------------------------------------------
% References verified against arXiv/publisher pages on 2026-09-04 (46 entries).
% The two Chen 2026 entries are concurrent preprints central to Sec. 5;
% re-check for any updated/published version immediately before submission.
\bibliographystyle{unsrtnat}
\bibliography{refs}

\appendix

\section{Claim-to-Artifact Map}
\label{app:map}

Every numbered claim in this paper is backed by a measurement report in \texttt{logs/} and by the scripts and result files that produced it in \texttt{experiments/}. Report names below are relative to \texttt{logs/}; artifact paths are relative to \texttt{experiments/}, both in the artifact repository linked above. Within an artifact cell, an entry that contains no slash is a file in the same directory as the first entry of that cell. Of the 32 reports released, the 24 cited here are the ones a claim in this paper depends on; the remaining eight cover the same corpus's supporting work (environment setup, PTQ deep-dive and raw run logs, QAT recovery, the BEVFormer/BEVDet capstone, and a CNN-detector accuracy axis) and are not load-bearing for any claim above. Claims are abbreviated here, and the absolute values they quote (latency, top-1, mAP, perf-per-watt) are batch-1, subset-based, and measured on different paths, so only the within-comparison \emph{relative} relations they support are valid (\S10).

\begingroup
\scriptsize
\setlength{\LTcapwidth}{\textwidth}
\begin{longtable}{L{0.9cm} L{4.0cm} L{4.3cm} L{5.0cm}}
\caption{Claim-to-artifact map. Each row names the report and the scripts behind one claim, so any number in the paper can be traced to the run that produced it.}\\
\toprule
\textbf{Sec.} & \textbf{Claim} & \textbf{Report (\texttt{logs/})} & \textbf{Scripts and results (\texttt{experiments/})} \\
\midrule
\endfirsthead
\multicolumn{4}{l}{\emph{Table \thetable, continued.}}\\
\toprule
\textbf{Sec.} & \textbf{Claim} & \textbf{Report (\texttt{logs/})} & \textbf{Scripts and results (\texttt{experiments/})} \\
\midrule
\endhead
\midrule
\multicolumn{4}{r}{\emph{continued on the next page}}\\
\endfoot
\bottomrule
\endlastfoot
\S4 (C1) & INT8 speedup sign set by the dot-product ISA, four CPUs, one model, one runtime & \texttt{stage4\_\allowbreak arm\_\allowbreak cpu\_\allowbreak fallback\_\allowbreak report.html} \newline \texttt{stage4\_\allowbreak imx8mn\_\allowbreak a53\_\allowbreak report.html} \newline \texttt{stage4\_\allowbreak jetson\_\allowbreak agx\_\allowbreak orin\_\allowbreak a78ae\_\allowbreak report.html} & \texttt{stage5\_\allowbreak infrastructure/\allowbreak cpu\_\allowbreak proxy/\allowbreak rpi\_\allowbreak bench.py} \newline \texttt{rpi\_\allowbreak bench\_\allowbreak lowmem.py} \newline \texttt{stage5\_\allowbreak infrastructure/\allowbreak cpu\_\allowbreak proxy/\allowbreak results/\allowbreak resnet50\_\allowbreak \_\allowbreak *.json} \\
\addlinespace[2pt]
\S5 (C2) & FP32 1000/1000 control; CPU$\leftrightarrow$CPU INT8 agreement 1000 / 965 / 961 / 958 & \texttt{stage4\_\allowbreak arm\_\allowbreak cpu\_\allowbreak fallback\_\allowbreak report.html} \newline \texttt{stage4\_\allowbreak imx8mn\_\allowbreak a53\_\allowbreak report.html} \newline \texttt{stage4\_\allowbreak jetson\_\allowbreak agx\_\allowbreak orin\_\allowbreak a78ae\_\allowbreak report.html} & \texttt{stage5\_\allowbreak infrastructure/\allowbreak cpu\_\allowbreak proxy/\allowbreak README.md} \newline \emph{(the agreement matrix)} \newline \texttt{stage5\_\allowbreak infrastructure/\allowbreak cpu\_\allowbreak proxy/\allowbreak raw/\allowbreak } \\
\addlinespace[2pt]
\S5 (C2) & CPU$\leftrightarrow$accelerator 961/1000 from the same QDQ artifact; iGPU INT8 top-1 0.7620 & \texttt{stage3\_\allowbreak jetson\_\allowbreak orin\_\allowbreak accuracy\_\allowbreak report.html} & \texttt{stage3\_\allowbreak tensorrt/\allowbreak jetson\_\allowbreak ondevice/\allowbreak accuracy/\allowbreak scripts/\allowbreak orin\_\allowbreak accuracy.py} \newline \texttt{analyze\_\allowbreak accuracy.py} \\
\addlinespace[2pt]
\S5 (C2\,$\ddagger$) & Vendor-NPU 939/1000, scales not held fixed (deployment observation only) & \texttt{stage4\_\allowbreak deepx\_\allowbreak dxm1\_\allowbreak accuracy\_\allowbreak report.html} & \texttt{stage4\_\allowbreak deepx\_\allowbreak dxm1/\allowbreak accuracy/\allowbreak scripts/\allowbreak npu\_\allowbreak infer.py} \newline \texttt{cpu\_\allowbreak infer.py} \newline \texttt{analyze\_\allowbreak dxm1\_\allowbreak acc.py} \\
\addlinespace[2pt]
\S5, \S11 & Power-of-two mitigation does not transfer across the x86$\leftrightarrow$A76 boundary: 958 $\to$ 869 (ceil) / 919 (nearest), gate NO-GO; cross-kernel $|\Delta\mathrm{logit}|$ $\times4.34$ / $\times2.35$, tie-breaking refuted; 30 runs bit-identical & \texttt{stage5\_\allowbreak pot\_\allowbreak scales\_\allowbreak report.html} & \texttt{stage5\_\allowbreak infrastructure/\allowbreak pot\_\allowbreak scales/\allowbreak pot\_\allowbreak rewrite.py} \newline \texttt{pot\_\allowbreak bench.py} \newline \texttt{pot\_\allowbreak agree.py} \newline \texttt{pot\_\allowbreak mech.py} \newline \texttt{stage5\_\allowbreak infrastructure/\allowbreak pot\_\allowbreak scales/\allowbreak results/\allowbreak } \\
\addlinespace[2pt]
\S6 (C3) & Qualcomm HTP silent BYO-QDQ failure 0.75 $\to$ 0.005; native path recovers 0.735 & \texttt{stage4\_\allowbreak qualcomm\_\allowbreak aihub\_\allowbreak report.html} & \texttt{stage4\_\allowbreak qualcomm\_\allowbreak aihub/\allowbreak scripts/\allowbreak qaihub\_\allowbreak int8.py} \newline \texttt{qaihub\_\allowbreak native\_\allowbreak quant.py} \newline \texttt{qaihub\_\allowbreak acc.py} \\
\addlinespace[2pt]
\S6 (C3) & DEEPX loud refusal (GraphStructureError, no engine); native path 0.7660 & \texttt{stage4\_\allowbreak deepx\_\allowbreak dxm1\_\allowbreak accuracy\_\allowbreak report.html} & \texttt{stage4\_\allowbreak deepx\_\allowbreak dxm1/\allowbreak accuracy/\allowbreak scripts/\allowbreak probe\_\allowbreak extqdq.py} \newline \texttt{compile\_\allowbreak dxm1.py} \\
\addlinespace[2pt]
\S7 (C4) & Three models, two regimes on one device; YOLOv5s 26.3$\times$ slower than ResNet-50 & \texttt{stage4\_\allowbreak deepx\_\allowbreak dxm1\_\allowbreak crossover\_\allowbreak report.html} & \texttt{stage4\_\allowbreak deepx\_\allowbreak dxm1/\allowbreak crossover/\allowbreak scripts/\allowbreak analyze\_\allowbreak profiler.py} \newline \texttt{build\_\allowbreak crossover\_\allowbreak summary.py} \\
\addlinespace[2pt]
\S7 (C4) & Fixed-compute output sweep, 12 points; transition band width = core count & \texttt{stage4\_\allowbreak deepx\_\allowbreak dxm1\_\allowbreak transition\_\allowbreak report.html} & \texttt{stage4\_\allowbreak deepx\_\allowbreak dxm1/\allowbreak transition/\allowbreak scripts/\allowbreak build\_\allowbreak transition\_\allowbreak models.py} \newline \texttt{run\_\allowbreak transition\_\allowbreak bench.sh} \newline \texttt{build\_\allowbreak transition\_\allowbreak summary.py} \newline \texttt{stage4\_\allowbreak deepx\_\allowbreak dxm1/\allowbreak transition/\allowbreak results/\allowbreak transition\_\allowbreak summary.json} \\
\addlinespace[2pt]
\S7 (C4) & Third regime: DETR host-CPU-compute-bound stage decomposition & \texttt{stage4\_\allowbreak deepx\_\allowbreak dxm1\_\allowbreak detr\_\allowbreak report.html} & \texttt{stage4\_\allowbreak deepx\_\allowbreak dxm1/\allowbreak detr/\allowbreak scripts/\allowbreak npu\_\allowbreak infer\_\allowbreak detr.py} \newline \texttt{analyze\_\allowbreak detr\_\allowbreak regime.py} \newline \texttt{stage4\_\allowbreak deepx\_\allowbreak dxm1/\allowbreak detr/\allowbreak results/\allowbreak detr\_\allowbreak dxm1\_\allowbreak summary.json} \\
\addlinespace[2pt]
\S7 & NPU vs.~host CPU in the compute-bound regime: $\times$11.42 throughput, $\times$29.29 perf/W & \texttt{stage4\_\allowbreak deepx\_\allowbreak dxm1\_\allowbreak report.html} & \texttt{stage4\_\allowbreak deepx\_\allowbreak dxm1/\allowbreak scripts/\allowbreak cpu\_\allowbreak bench.py} \newline \texttt{analyze\_\allowbreak profiler.py} \newline \texttt{build\_\allowbreak summary.py} \\
\addlinespace[2pt]
\S8.1 & DETR INT8 collapse $-$42.9\%; op-selection no-op; two-way ablation & \texttt{stage2\_\allowbreak detr\_\allowbreak quantization\_\allowbreak report.html} & \texttt{stage2\_\allowbreak detr/\allowbreak s2\_\allowbreak 04\_\allowbreak ptq.py} \newline \texttt{s2\_\allowbreak 07\_\allowbreak coco\_\allowbreak eval.py} \newline \texttt{s2\_\allowbreak 08\_\allowbreak quantize\_\allowbreak mixed.py} \newline \texttt{s2\_\allowbreak 09\_\allowbreak quantize\_\allowbreak ablation.py} \\
\addlinespace[2pt]
\S8.1 & Jetson symmetric-requantization cross-confirmation $-$43.8\% & \texttt{stage3\_\allowbreak jetson\_\allowbreak orin\_\allowbreak detr\_\allowbreak accuracy\_\allowbreak report.html} & \texttt{stage3\_\allowbreak tensorrt/\allowbreak jetson\_\allowbreak ondevice/\allowbreak detr\_\allowbreak accuracy/\allowbreak scripts/\allowbreak detr\_\allowbreak sym\_\allowbreak export.py} \newline \texttt{orin\_\allowbreak detr\_\allowbreak map.py} \\
\addlinespace[2pt]
\S8.1 & SmoothQuant recovers 59.9\% of the gap (torch fake-quant path) & \texttt{stage2\_\allowbreak smoothquant\_\allowbreak report.html} & \texttt{stage2\_\allowbreak smoothquant/\allowbreak sq\_\allowbreak 01\_\allowbreak modelopt\_\allowbreak api.py} \newline \texttt{sq\_\allowbreak 03\_\allowbreak absmax\_\allowbreak smooth.py} \newline \texttt{sq\_\allowbreak 04\_\allowbreak alpha\_\allowbreak sweep.py} \\
\addlinespace[2pt]
\S8.1 & SmoothQuant recovers only ${\sim}$9\% on-device (Gemm-only INT8 coverage) & \texttt{stage3\_\allowbreak jetson\_\allowbreak orin\_\allowbreak detr\_\allowbreak smoothquant\_\allowbreak report.html} & \texttt{stage3\_\allowbreak tensorrt/\allowbreak jetson\_\allowbreak ondevice/\allowbreak detr\_\allowbreak smoothquant/\allowbreak scripts/\allowbreak detr\_\allowbreak sq\_\allowbreak export.py} \newline \texttt{orin\_\allowbreak detr\_\allowbreak sq\_\allowbreak map.py} \\
\addlinespace[2pt]
\S8.2--\S8.3 & DEEPX auto-split (no collapse), 23.99~MB handoff; two-quantizer comparison & \texttt{stage4\_\allowbreak deepx\_\allowbreak dxm1\_\allowbreak detr\_\allowbreak report.html} & \texttt{stage4\_\allowbreak deepx\_\allowbreak dxm1/\allowbreak detr/\allowbreak scripts/\allowbreak analyze\_\allowbreak detr\_\allowbreak map.py} \newline \texttt{build\_\allowbreak detr\_\allowbreak summary.py} \\
\addlinespace[2pt]
\S8 (note) & NVDLA INT8-only datapath, 51.29~inf/s/W; DETR DLA fragmentation (16 ForeignNodes) & \texttt{stage3\_\allowbreak jetson\_\allowbreak orin\_\allowbreak ondevice\_\allowbreak report.html} \newline \texttt{stage3\_\allowbreak jetson\_\allowbreak orin\_\allowbreak concurrent\_\allowbreak power\_\allowbreak report.html} \newline \texttt{stage3\_\allowbreak jetson\_\allowbreak orin\_\allowbreak detr\_\allowbreak report.html} & \texttt{stage3\_\allowbreak tensorrt/\allowbreak jetson\_\allowbreak ondevice/\allowbreak scripts/\allowbreak ppw.py} \newline \texttt{concurrent.py} \newline \texttt{power\_\allowbreak sweep.py} \newline \texttt{stage3\_\allowbreak tensorrt/\allowbreak jetson\_\allowbreak ondevice/\allowbreak detr/\allowbreak scripts/\allowbreak detr\_\allowbreak bench.py} \\
\addlinespace[2pt]
\S9 (C8) & Subset inflation +9.77~pp (mean of 8 configs); preprocessing $-$1.07~pp & \texttt{stage1\_\allowbreak 50k\_\allowbreak rerun\_\allowbreak reproduction\_\allowbreak report.html} \newline \texttt{stage1\_\allowbreak real\_\allowbreak imagenet\_\allowbreak report.html} & \emph{(no separate script dir; procedure is in the report and in study\_guide/03, /10)} \\
\addlinespace[2pt]
\S9 (C8) & Per-layer SQNR does not predict $\Delta$top-1 (Spearman $\rho$ = $-$0.030, 21 layers) & \texttt{stage1\_\allowbreak 50k\_\allowbreak rerun\_\allowbreak reproduction\_\allowbreak report.html} & \emph{(no separate script dir; procedure is in the report and in study\_guide/03)} \\
\addlinespace[2pt]
\S9 (C8) & TensorRT EP listed but silently on CPU, p50 11.83 $\to$ 0.41~ms once fixed & \texttt{stage0.5\_\allowbreak ladder\_\allowbreak log.html} & \emph{(no separate script dir; fix and probe are in study\_guide/01)} \\
\addlinespace[2pt]
\S9 (C8) & Zero-copy output-buffer aliasing collapses top-1 to 0.0014, no error raised & \texttt{stage5\_\allowbreak infrastructure\_\allowbreak report.html} & \texttt{stage5\_\allowbreak infrastructure/\allowbreak bench/\allowbreak run\_\allowbreak bench.py} \newline \texttt{stage5\_\allowbreak infrastructure/\allowbreak bench/\allowbreak report/\allowbreak generate.py} \newline \texttt{stage5\_\allowbreak infrastructure/\allowbreak bench/\allowbreak tests/\allowbreak test\_\allowbreak regression.py} \\
\addlinespace[2pt]
\S9 (C8) & Opset down-convert ``succeeds'' (exit 0) while emitting an invalid graph & \texttt{stage1\_\allowbreak quantization\_\allowbreak log.html} & \emph{(no separate script dir; procedure is in the report)} \\
\addlinespace[2pt]
\S9 (C8) & pip TensorRT wheel ships without the `trtexec' binary the tutorials assume & \texttt{stage3\_\allowbreak tensorrt\_\allowbreak report.html} & \texttt{stage3\_\allowbreak tensorrt/\allowbreak t01\_\allowbreak env.py} \newline \texttt{t02\_\allowbreak latency\_\allowbreak 3point.py} \\
\end{longtable}
\endgroup

\end{document}